\documentstyle[twocolumn,aps,epsfig]{revtex}
\begin{document}
\draft
\title{Photon Polarization Measurements without the Quantum Zeno Effect}

\author{V. Kidambi, A. Widom, and C. Lerner}
\address{Physics Department, Northeastern University, Boston MA 02115}
\author{Y.N. Srivastava }
\address{Physics Department \& INFN, University of Perugia, Perugia, Italy}
\maketitle

\begin{abstract}
We consider a photon beam incident on a stack of polarizers 
as an example of a von Neumann projective measurement, theoretically   
leading to the quantum Zeno effect. The Maxwell theory     
(which is equivalent to the single photon Schr\"odinger equation) 
describes measured polarization phenomena, but without recourse to 
the notion of a projective measurement. 
\end{abstract}  

\pacs{PACS: 03.65 -w, 03.65 -Bz, 03.65 -Pm.}  
\narrowtext

\section{Introduction}

The notion of a projective quantum measurement was introduced by 
von Neumann\cite{1} in his treatise on the mathematical foundations 
of quantum mechanics. The idea is the following: (i) Every quantum 
measurement yielding the data ``yes'' or ``no'' to an experimental question 
is described by the projection operators ${\cal P}={\cal P}^\dagger=
{\cal P}^2$ (for ``yes'') or ${\cal Q}={\cal Q}^\dagger={\cal Q}^2$ 
(for ``no''),  
\begin{equation}
{\cal P}+{\cal Q}=1.
\end{equation}
(ii) If the initial state of a quantum object {\em before} a  
measurement is $|\Psi_{before}>$, and if the measurement yields the 
experimental answer ``yes'' to a question ${\cal P}$, then the state 
of the quantum object {\em after} the measurement $|\Psi_{after}>$ 
is constructed according to the {\em collapse} of the quantum state 
rule 
\begin{equation}
\big|\Psi_{before}\big>\to \big|\Psi_{after}\big>=\ 
{{\cal P}\big|\Psi_{before} \big>\over 
\ \sqrt{\big<\Psi_{before}\big|{\cal P}\big|\Psi_{before}\big>}\ }\ .
\end{equation}
Eq.(2) is part of a calculation scheme which is {\em not} a unitary 
development of the quantum state {\em during a measurement}. The unitary 
behavior (implicit in the Schr\"odinger equation for a quantum object) 
was thought by von Neumann to hold only {\em between measurements}\cite{2}.

To see how this works, consider a beam of photons which has passed through 
an ``upwards'' polarizer as in Figs.1(a) and (b). If a second polarizer  
has an axis at an angle of $\theta $ with respect to the upward 
direction, then one writes the photon wave function $|\uparrow >$ in the 
basis  
\begin{equation}
\big|\uparrow \big>=\cos \theta \big|\nwarrow \big>+
\sin \theta \big|\nearrow \big>.
\end{equation} 
The probability to pass through the second polarizer is then 
\begin{equation}
p(\theta )=\big|\big<\nwarrow \big|\uparrow \big>\big|^2
=\cos^2\theta .
\end{equation}

The cases shown in Fig.1 below work as follows: In Fig.1(a), the second 
polarizer is set orthogonal to the first polarizer so that 
no photons will pass; i.e. 
\begin{equation}
p_a(\pi /2 )=\big|\big<\leftarrow \big|\uparrow \big>\big|^2
=\cos^2 (\pi /2)=0.
\end{equation} 
This probability can be increased, as in Fig.1(b) below, by placing an 
intermediate polarizer between the first and last orthogonal polarizers.

\begin{figure}[htbp]
\begin{center}
\mbox{\epsfig{file=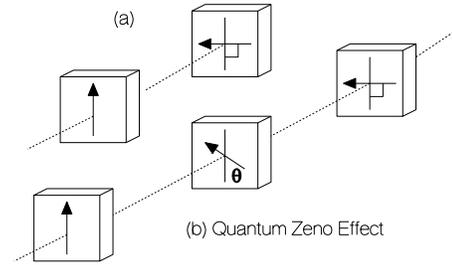,height=70mm}}
\caption{A photon beam polarized ``upwards'' will not pass through a second 
orthogonal polarizer as shown in (a). If an intermediate polarizer, as shown 
in (b), is placed before the final polarizer, then some photons in 
the beam will pass through the final polarizer. The intermediate 
measurement is then thought to induce the quantum Zeno effect.}
\label{zfig1}
\end{center}
\end{figure}

In Fig.1(b), a fraction $\cos^2 \theta $ of the $|\uparrow >$ 
photon beam will pass through the intermediate polarizer and 
{\em collapse into the state $|\nwarrow >$}. Then, a fraction 
$\cos^2 \big((\pi /2)-\theta \big)$ of the $|\nwarrow >$ photon beam 
will pass through the final polarizer and again {\em collapse into the state 
$|\leftarrow >$}. The above two {\em collapse of the quantum state 
processes} dictate that one multiply probabilities. The final probability 
that a photon in the upward polarized beam passes through both the 
intermediate and final polarizers is then given by 
\begin{equation}
p_b(\theta )=\cos^2 \theta \cos^2 \big((\pi /2)-\theta \big)=
(1/4)\sin^2 (2\theta ).
\end{equation} 
That an intermediate measurement can {\em increase} the probability of 
passing through a polarizer stack, say $(1/4)=p_b(\pi /4)>p_a(\pi /2)=0$, 
is an example of what is thought to be a quantum Zeno effect
\cite{3,4,5,6,7,8,9,10,11,12,13}

A textbook quantum Zeno effect problem in quantum mechanics\cite{14} is 
shown in Fig.2.

\begin{figure}[htbp]
\begin{center}
\mbox{\epsfig{file=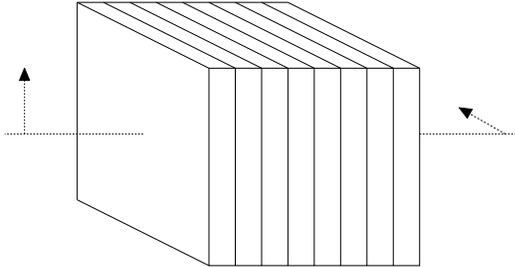,height=70mm}}
\caption{A vertical polarized photon beam enters the stack of N=8 polarizers, 
and a horizontally polarized photon beam leaves the stack.}
\label{zfig2}
\end{center}
\end{figure}

A beam of photons in state $|\uparrow >$ is incident on a stack of $N$ 
polarizers  each of which has a polarization axis rotated 
by an angle of $\Delta \theta_N =(\pi /2N)$ relative to the previous 
polarizer. According to the von Neumann projection postulate\cite{15}, 
the quantum state of each photon will {\em collapse} $N$ times, yielding 
a probability 
\begin{equation}
P_N=\big(\cos^{2}(\Delta \theta_N )\big)^N=
\Big\{\cos^{2}\Big({\pi \over 2N}\Big)\Big\}^N  
\end{equation}
of passing through the total stack as pictured in Fig.2. Note, in the 
formal limit of an {\em infinite number of quantum state collapse events},
\begin{equation}
\lim_{N\to \infty }P_N =1, \ \ \ {\rm (Quantum\ Zeno\ Effect)}.
\end{equation} 
Theoretically {\it every photon passes through the stack}. In real 
experiments, there is always some attenuation of a photon beam 
passing through many polarizers. 

Although the large $N$ quantum Zeno effect does appear to be verified 
in the laboratory\cite{16}, we have {\em a very uneasy feeling} about 
attributing such observations to the rule of {\em collapsing quantum 
states}. After all, long before von Neumann considered projective 
measurements, Maxwell had developed a reliable electromagnetic radiation 
theory which perfectly well determined how an electromagnetic wave will 
(or will not) pass through stacks of polarizers. {\em And not even once} 
did Maxwell feel obliged to discuss the collapse of the electromagnetic 
radiation field functions. For that matter, neither Bohr nor 
Einstein\cite{17} had {\em ever} felt obliged to discuss the so-called wave 
function collapse. 

Our purpose is to discuss why the projection postulate (collapse of the 
quantum state), {\em certainly may and probably should} 
be eliminated from ones thinking in quantum mechanics. 
The application of this view towards photon polarization measurements 
is typical of a more general state of affairs. If one returns 
to the original Copenhagen interpretation of the quantum mechanical 
formalism, then the quantum state describes an {\em ensemble of experiments}. 
Each photon certainly does {\em not} carry its own wave function which 
might then undergo a collapse whenever a photon disappears (e.g. whenever 
a photon is absorbed in a polarizer). The wave function (in reality) 
{\em describes an ensemble} of many photons. 

The collapse of the quantum state whenever a photon is absorbed 
closely resembles the collapse of a probability table (at the 
automobile insurance company) every time a car crashes. Actually, 
the probability table at the insurance company stays intact 
during a car crash. It is the {\em automobile} and {\em not} 
the probability that collapses. This constitutes an altogether 
{\em different} pile of scrap metal, even if the cross section 
for two automobiles to scatter off one another were computed 
with a quantum mechanical amplitudes to include the small 
diffraction effects.

In Sec.II, it will be shown how the Schr\"odinger equation for photons in 
the vacuum is equivalent to the vacuum Maxwell field equations. In 
Sec.III, the  Schr\"odinger equation for photons propagating in 
continuous media (e.g. polarizers) will be explored. The photon wave 
function follows from the Maxwell electromagnetic theory in continuous media. 
However, the electromagnetic wave attenuates if the continuous 
media exhibits dissipation in the form of finite conductivity. 
The complete quantum system (polarizers plus photons) 
obeys {\em a big monster} Schr\"odinger equation with a huge 
number of degrees of freedom and 
{\it does not at all require the projective measurement} 
formalism of Eq.(2). This feature of quantum mechanics 
is valid for many systems as will be discussed in Sec.IV.  

In Sec.V we compute, {\em for a realistic model of polarizers}, the 
Maxwell equation formulation of the Zeno effect. The {\em classical Maxwell 
wave} computation (without projection operators) yields the same result 
as {\em the quantum photon Schr\"odinger amplitude} (without projection 
operators). 

In an appropriate regime, the projection postulate does indeed 
give an accurate answer, but this may only be proved by solving 
the full Maxwell theory without the projection postulate. One might be 
led to ask why should one {\em ever} add to the quantum theory a 
projection theory of measurements. A hint as to 
why the projection postulate is so popular might be connected to the fact 
that the Maxwell theory computation (to be presented) required a fast 
computer. More than mere speed, we required algorithms sufficient to 
maintain $\sim 10^3$ significant figure accuracy during the entire 
calculation of the photon transmission probability. Without recently 
developed precise numerical algorithms, one might be coerced into 
employing the projection postulate. 

In the concluding Sec.VI, we question the utility of replacing the wisdom 
of experimentalists and engineers who design and build laboratory 
equipment with a {\em merely} philosophical view of projective  
measurements. It is somewhat unreasonable to represent real 
laboratory equipment with mathematical projection operators. 

\section{Photon Wave Functions}

For a spin $S$ particle\cite{18} one requires three Hermitian matrices 
${\bf S}=(S_1,S_2,S_3)$ obeying 
\begin{equation}
[S_i,S_j]=i\epsilon_{ijk}S_k, \ \ 
S_1^2+S_2^2+S_3^2=S(S+1).
\end{equation}
The photon is a spin one ($S=1$) particle so we may choose the spin 
matrices
\begin{equation}
S_1=\pmatrix{ 0 & 0 & 0\cr 
0 & 0 & -i\cr 
0 & i & 0},\ \  
S_2=\pmatrix{ 0 & 0 & i\cr 
0 & 0 & 0\cr 
-i & 0 & 0},
\end{equation}
and
\begin{equation}
S_3=\pmatrix{ 0 & -i & 0\cr 
i & 0 & 0\cr 
0 & 0 & 0}. 
\end{equation}
With ${\bf p}=-i\hbar {\bf \nabla}$, the vacuum Schr\"odinger equation for 
free photons reads 
\begin{equation}
i\hbar \Big({\partial \psi ({\bf r},t)\over \partial t}\Big)
=c\big({\bf S\cdot p} \big)\psi ({\bf r},t).
\end{equation}
Writing 
\begin{equation}
\psi ({\bf r},t)=\pmatrix{F_1({\bf r},t) \cr 
F_2({\bf r},t) \cr F_3({\bf r},t) },
\end{equation}
allows one to cast the photon Schr\"odinger Eq.(12) into the (perhaps) 
more simple vector form 
\begin{equation}
i\Big({\partial {\bf F} ({\bf r},t)\over \partial t}\Big)=
c\ {\bf curl\  F} ({\bf r},t).
\end{equation}
The vector field ${\bf F}({\bf r},t)$ should be constrained to obey 
\begin{equation}
div{\bf F}({\bf r},t)=0.
\end{equation}
Eq.(15) is equivalent to only allowing two independent spin states 
(say superpositions of only two transverse polarizations), 
thereby {\em eliminating} a third (longitudinal polarized) 
``zero energy'' mode. Finally, Eqs.(14) and (15) take on a clearly 
recognizable form if one 
defines the real and imaginary parts of ${\bf F}({\bf r},t)$ to be, 
respectively, the electric field ${\bf E}({\bf r},t)$ and the 
magnetic field ${\bf B}({\bf r},t)$\cite{19}\cite{20}\cite{21}; i.e.
\begin{equation}
{\bf F}({\bf r},t)={\bf E}({\bf r},t)+i{\bf B}({\bf r},t).
\end{equation}
The photon Schr\"odinger Eq.(12) then appears equivalent to the vacuum 
Maxwell theory; i.e.  
\begin{equation}
\Big({\partial {\bf E} ({\bf r},t)\over \partial t}\Big)=
c\ {\bf curl\ B}({\bf r},t),
\end{equation}
and 
\begin{equation}  
\Big({\partial {\bf B} ({\bf r},t)\over \partial t}\Big)=
-c\ {\bf curl\ E}({\bf r},t),
\end{equation}
with the vacuum Gauss law Eq.(15) constraints 
\begin{equation}
div {\bf B}({\bf r},t)=0,\ \ \ 
div {\bf E}({\bf r},t)=0. 
\end{equation}

It is of interest to discuss the relativistic Lorentz symmetry of 
the photon Schr\"odinger Eq.(12). The proof of Lorentz symmetry is 
actually well known since we have already proved equivalence with the
Maxwell theory. However, the discrete Lorentz symmetries are still 
worthy of further discussion\cite{22,23,24,25,26,27,28}. If we 
consider the classical Lorentz force on a charge equation 
${\bf f}=e({\bf E}+({\bf v\times B})/c)$, 
and then allow the charge to change sign $e\to -e$, then the 
force ${\bf f}$ will remain the same if we also allow 
${\bf E}\to -{\bf E}$ and ${\bf B}\to -{\bf B}$. Thus, the electromagnetic 
field Eq.(16) is odd under charge conjugation,
\begin{equation} 
\hat{C}{\bf F}=-{\bf F}.
\end{equation}
Under time reversal, ${\bf E}\to {\bf E}$ and ${\bf B}\to -{\bf B}$ so 
that  
\begin{equation}  
\hat{T}{\bf F}={\bf F}^*.
\end{equation}
Finally, under parity ${\bf E}\to -{\bf E}$ and ${\bf B}\to {\bf B}$ so 
that  
\begin{equation}
\hat{P}{\bf F}=-{\bf F}^*.
\end{equation}
Note  
\begin{equation}
\hat{T}\hat{C}\hat{P}=1,
\end{equation}
as would be expected from relativistic wave equations. 

\section{Photon Wave Functions in Continuous Media}

The Maxwell theory for an electromagnetic disturbance moving through 
continuous media\cite{29,30,31,32,33,34} may be formulated starting 
from 
\begin{equation}
div{\bf E}({\bf r},t)=4\pi \rho ({\bf r},t),\ \ \ 
div{\bf B}({\bf r},t)=0,
\end{equation}
and 
\begin{equation}
\Big({\partial {\bf E} ({\bf r},t)\over \partial t}\Big)=
c\ {\bf curl\  B} ({\bf r},t)-4\pi {\bf J}({\bf r},t),
\end{equation}
\begin{equation}
\Big({\partial {\bf B} ({\bf r},t)\over \partial t}\Big)=
-c\ {\bf curl\  E} ({\bf r},t).
\end{equation}
It is sufficiently general to supplement Eqs.(24), (25) and (26) by 
employing a linear causal relationship between the current and the electric 
field, i.e. 
\begin{equation}
{\bf J}({\bf r},t)=\int_0^\infty \int 
{\bf K}({\bf r},{\bf r}',s){\bf \cdot E}({\bf r}',t-s)
d^3 {\bf r}'ds.
\end{equation}
If, for a complex frequency $\zeta $ where $\Im m\ \zeta >0$, 
\begin{equation}
{\bf E} ({\bf r},t)=\Re e \big\{{\bf E}_0 ({\bf r},\zeta)
e^{-i\zeta t}\big\},
\end{equation}
\begin{equation}
{\bf J} ({\bf r},t)=\Re e \big\{{\bf J}_0 ({\bf r},\zeta )
e^{-i\zeta t}\big\}, 
\end{equation}
Eqs.(27), (28) and (29) serve to define the non-local conductivity 
${\bf \sigma }({\bf r},{\bf r}',\zeta )$, i.e.  
\begin{equation}
{\bf J}_0 ({\bf r},\zeta )=\int {\bf \sigma }({\bf r},{\bf r}',\zeta )
{\bf \cdot E}_0({\bf r}',\zeta)d^3{\bf r}',
\end{equation}
where  
\begin{equation}
{\bf \sigma }({\bf r},{\bf r}',\zeta )=\int_0^\infty e^{i\zeta s}
{\bf K}({\bf r},{\bf r}',s)ds.
\end{equation}
Some workers prefer to complete the Maxwell equations in continuous media 
by employing a constitutive equation relating the Maxwell displacement 
field ${\bf D}({\bf r},t)$ to the electric field ${\bf E}({\bf r},t)$. 
In the frequency domain, the dielectric response 
${\bf \epsilon }({\bf r},{\bf r}',\zeta )$ of the continuous media is defined 
as  
\begin{equation}
{\bf D}_0 ({\bf r},\zeta )=\int {\bf \epsilon }({\bf r},{\bf r}',\zeta )
{\bf \cdot E}_0({\bf r}',\zeta)d^3{\bf r}'.
\end{equation}
Just so long as the various response functions are maintained as non-local 
(in space and time), which of the many equivalent response functions are 
actually used in a calculation remains a matter of convenience. 
For example, Eqs.(30) and (32) are equivalent if we choose 
\begin{equation}
{\bf \epsilon }({\bf r},{\bf r}',\zeta )=
{\bf 1}\delta ({\bf r}-{\bf r}')+
\Big({4\pi i {\bf \sigma }({\bf r},{\bf r}',\zeta )\over \zeta }\Big).
\end{equation}

The dielectric, magnetic, or conductivity response functions conventionally 
used to describe continuous media now make an appearance into the 
Schr\"odinger equation for a photon moving through the continuous media 
via the photon ``self-energy'' insertion. To see what is involved, 
we may write the Maxwell Eqs.(24) and (25) as 
\begin{equation}
div{\bf F} ({\bf r},t)=4\pi \rho ({\bf r},t),
\end{equation}
and 
\begin{equation} 
i\Big({\partial {\bf F} ({\bf r},t)\over \partial t}\Big)=
c\ {\bf curl\  F} ({\bf r},t)-4\pi i{\bf J}({\bf r},t).
\end{equation} 
where Eq.(16) has been invoked. Employing Eqs.(21), (27) and 
(35), we find our central result for the photon Schr\"odinger equation 
in continuous media 
$$ 
i\Big({\partial {\bf F} ({\bf r},t)\over \partial t}\Big)=
c\ {\bf curl\  F} ({\bf r},t)+
$$
\begin{equation} 
\int_0^\infty \int {\bf \Sigma }({\bf r},{\bf r}',s)
{\bf F} ({\bf r}',t-s)d^3 {\bf r}'ds.
\end{equation} 
where the photon self-energy part is defined as 
\begin{equation}
{\bf \Sigma }({\bf r},{\bf r}',s)=-2\pi i
{\bf K}({\bf r},{\bf r}',s)(1+\hat{T}).
\end{equation} 
Using Eqs.(13) and (36), one may write the photon Schr\"odinger 
equation in the form 
$$ 
i\hbar\Big({\partial \psi ({\bf r},t)\over \partial t}\Big)=
c\big({\bf S\cdot p}\big) \psi ({\bf r},t)+
$$
\begin{equation} 
\hbar\int_0^\infty \int {\bf \Sigma }({\bf r},{\bf r}',s)
\psi ({\bf r}',t-s)d^3 {\bf r}'ds.
\end{equation} 

In principle the calculation of the self energy part 
${\bf \Sigma }({\bf r},{\bf r}',s)$ is equivalent to the 
calculation of the conductivity 
${\bf \sigma }({\bf r},{\bf r}',\zeta )$
as is evident from Eqs.(21), (31), and (37). In a practical 
calculations involving the photon Schr\"odinger equation 
in a polarizer, one should use the conductivity and dielectric      
response functions as those of the optical engineers who designed 
the device\cite{35,36,37,38,39}. Finally, the non-local form of 
Eq.(38) is shared by many other quantum mechanical systems as we 
shall now briefly explore.

\section{Non-Local Schr\"odinger Equations}

Although we do not adhere to the non-unitary collapse of the 
quantum state Eq.(2) as postulated by von Neumann, we do find the 
notion of projection operators (also introduced into quantum mechanics 
by von Neumann) to be more than just convenient. Eq.(1) will thereby 
be invoked in what follows. For a macroscopic 
laboratory quantum system we here insist that a {\em monster} 
Schr\"odinger equation with a huge number of degrees of freedom still 
holds true,
\begin{equation} 
i\hbar \Big({\partial \over \partial t}\Big)|\Psi (t)>
={\cal H}_{tot}|\Psi (t)>.
\end{equation}
We do not contemplate the {\em ultimate monster wave function} of the {\em 
whole universe}, not only because we are not smart enough to find it, but 
because we do not know how to build an experimental 
{\em ensemble of universes} to test a hypothetical quantum state of 
{\em our universe}. We leave such considerations to people more wise  
than are we.

We are happy to say something meaningful about a very small physical 
piece of a laboratory quantum state; e.g. a ``projected piece''   
\begin{equation} 
\big|\psi (t)\big>={\cal P}\big|\Psi (t)\big>.
\end{equation}
We leave the rest of the monster wave function,  
\begin{equation} 
\big|\phi (t)\big>={\cal Q}\big|\Psi (t)\big>,
\end{equation}
to the engineers who design the experiment. With the little piece of the wave 
function Eq.(40) and the big piece of the wave function Eq.(41), we rewrite 
Eq.(39) as 
\begin{equation}
i\hbar \Big({\partial \over \partial t}\Big)
\pmatrix{|\psi (t)> \cr |\phi (t)>}=
\pmatrix{ H & V\cr V^\dagger & H'}
\pmatrix{|\psi (t)> \cr |\phi (t)>},
\end{equation}
where $H={\cal P}{\cal H}_{tot}{\cal P}$, 
$V={\cal P}{\cal H}_{tot}{\cal Q}$, 
$V^\dagger={\cal Q}{\cal H}_{tot}{\cal P}$ and 
$H'={\cal Q}{\cal H}_{tot}{\cal Q}$.

We can only hope to find $|\psi (t)>$ which obeys 
\begin{equation}
i\hbar \Big({\partial \over \partial t}\Big)\big| \psi (t) \big>
=H\big| \psi (t) \big>+V\big| \phi (t) \big>,
\end{equation}
so we solve for the monster piece 
\begin{equation}
i\hbar \Big({\partial \over \partial t}\Big)\big| \phi (t) \big>
=H'\big| \phi (t) \big>+V^\dagger \big| \psi (t) \big>,
\end{equation}
subject to the causal boundary condition 
\begin{equation}
\big| \phi (t) \big>=-\Big({i\over \hbar}\Big)\int_0^\infty  
e^{-iH's/\hbar}V^\dagger\big|\psi(t-s)\big>ds.
\end{equation}

From Eqs.(43) and (45) we find the general non-local Schr\"odinger 
equation for the piece of the wave function that is of interest; 
It is  
\begin{equation}
i\hbar \Big({\partial \over \partial t}\Big)\big|\psi (t)\big>
=H\big|\psi (t) \big>+
\hbar \int_0^\infty \Sigma (s)\big|\psi(t-s)\big>ds,
\end{equation}
where 
\begin{equation}
\Sigma (s)=
-\Big({i\over \hbar^2}\Big)Ve^{-iH's/\hbar }V^\dagger .
\end{equation}

The full monster Schr\"odinger Eq.(39) for complete wave function 
$|\Psi (t)>$ reduces to the non-local in time  Schr\"odinger Eq.(46) 
for the small subspace piece of the wave function  $|\psi(t)>$. 
A particular example of Eq.(46) is the photon Schr\"odinger Eq.(38). 
The point is that when a photon propagates inside of a continuous media 
device, say a polarizer, the {\em monster wave function} 
includes {\em both} the photon {\em and} the polarizer degrees of freedom. 
In a one photon projected subspace, the quantum state of the polarizer  
is fixed. In addition to the polarizer, there is but 
one propagating photon. If the photon is absorbed, then an electronic 
excitation is produced which heats up the polarizer. The piece of 
the monster quantum state representing this absorbed photon situation 
is not contained in the single photon subspace. The single photon 
subspace wave function is then attenuated due to photon absorption 
in the polarizer. All of this (including the attenuation of the Maxwell 
wave in the medium) is described by the one photon Schr\"odinger equation, 
{\it without} recourse to the projection postulate. Maxwell was so 
lucky to be able to write down the Schr\"odinger equation for the photon 
in matter, including attenuation, {\em without} knowing anything about 
quantum states (be they monster or otherwise).

It is of interest to look for a fixed energy solution of the subspace 
wave function Eq.(46) of the form 
\begin{equation}
\big|\psi (t)\big>=\big|\psi_E\big>e^{-iEt/\hbar}.    
\end{equation}
Eqs.(46), (47) and (48) imply 
\begin{equation}
\Big\{H+\Delta (E)-\Big({i\hbar \over 2}\Big)\Gamma (E)\Big\}
\big|\psi_E\big>=E\big|\psi_E\big>,
\end{equation}
where 
\begin{equation}
\Delta (E)-\Big({i\hbar \over 2}\Big)\Gamma (E)=
V\Big({1\over E-H'+i0^+}\Big)V^\dagger .
\end{equation}
That the effective energy dependent Hamiltonian on the left hand side of 
Eq.(49)  
\begin{equation}
{\cal H}_{eff}(E)=H+\Delta(E)-\Big({i\hbar \over 2}\Big)\Gamma (E)
\end{equation}
is not Hermitian, is due to the ``Fermi platinum rule'' transition rates 
to leave the small subspace and wander into the remaining monster subspaces. 
The Fermi platinum rule  
\begin{equation} 
\Gamma (E)=\Big({2\pi \over \hbar}\Big)V\delta (E-H')V^\dagger  
\end{equation}
is just a little bit more valuable than the more usual 
perturbative Fermi golden rule, since Eq.(52) is rigorously true to 
all orders of perturbation theory. The fixed energy $E$ case 
corresponds to having a fixed photon frequency $\omega =(E/\hbar)$ 
for the electromagnetic problem at hand. Let us return to this problem. 

\section{Maxwell Theory and Polarizers}

The Maxwell theory of a transverse wave moving through a polarizer may 
be formulated following the discussion of Born and Wolf\cite{40}. 
(i) If a transverse wave propagates in the $z$-direction, then the wave 
may be described by four electromagnetic field components, say  
\begin{equation}
\chi(z)=\pmatrix{E_1(z)\cr E_2(z)\cr B_1(z)\cr B_2(z)}.
\end{equation}
If the polarizer is oriented by a rotation angle $\theta $, 
as in Fig.1(b), then the dielectric response tensor will have the form 
\begin{equation}
\epsilon(\theta )=
\left({\tilde{\epsilon}_1+\tilde{\epsilon}_2\over 2}\right)+
\left({\tilde{\epsilon}_1-\tilde{\epsilon}_2\over 2}\right)
e^{-i\tau_2\theta }\tau_3e^{i\tau_2\theta },
\end{equation}
where we have employed the $2\times 2$ Pauli matrices 
\begin{equation}
\tau_1=\pmatrix{0 & 1 \cr 1 & 0},\ \ \   
\tau_2=\pmatrix{0 & -i \cr i & \ 0},  
\end{equation}
and
\begin{equation}
\tau_3=\pmatrix{1 & \ 0 \cr 0 & -1}.
\end{equation}
The complex principal eigenvalues of the dielectric tensor at frequency 
$\omega $ have the form 
\begin{equation}
\tilde{\epsilon}_j=\varepsilon_j+
\left(4\pi i\sigma_j\over \omega \right),\ \ \  
j=1,2\ .
\end{equation}
The real parts of the dielectric eigenvalues $\varepsilon_{1,2}$ determine 
the velocities of the waves polarized in the principal directions. The 
conductivities $\sigma_{1,2}$ determine the dissipative attenuation of the 
waves polarized in the principle directions. For a well designed polarizer, 
a wave with an allowed polarization direction will propagate unattenuated 
while a wave with a disallowed polarization will be strongly attenuated. 

{\em It is the anisotropic dissipative conductivity that is often modeled 
by the quantum projective postulate for the case of photon polarization}. 
On the other hand, conductivity is evidently {\it classical} since both 
Ohm and Maxwell were dead and gone {\em before} quantum mechanics had been 
invented.

For the classical Maxwell calculation, the wave at the beginning 
$(z=0)$ of the polarizer is related to the wave at the end $(z=a)$ 
of the polarizer by a $4\times 4$ transfer matrix
\begin{equation} 
\chi (a)={\cal M}(\theta ,\xi )\chi (0), \ \ \   
\xi =\Big({\omega a\over c}\Big), 
\end{equation}
$\chi(z)$ is defined in Eq.(53), and in partitioned matrix form  
$$
{\cal M}(\theta ,\xi )=
$$
\begin{equation}
\pmatrix{ \cos(\xi \sqrt{\epsilon(\theta )}) & 
-\big({\sin(\xi \sqrt{\epsilon(\theta )})/ 
\sqrt{\epsilon(\theta )} }\big)\tau_2 \cr
\tau_2 \sqrt{\epsilon(\theta ) }\sin(\xi \sqrt{\epsilon(\theta )}) &
\tau_2\cos(\xi \sqrt{\epsilon (\theta )})\tau_2 }. 
\end{equation}
For a stack of $N$ polarizers, each of width $a$ and oriented at angles 
$(\theta_1,...,\theta_N)$, one may repeat the above process going from 
the beginning $(z=0)$ of the stack to the end $(z=Na)$ of the stack
\begin{equation}
\chi (Na)={\cal M}_{N}\chi (0), \ \   
\xi =\Big({\omega a\over c}\Big).
\end{equation}
One Multiplies the transfer matrices for each polarizer in the stack 
\begin{equation}
{\cal M}_{tot,N}={\cal M}(\theta_N,\xi)...
{\cal M}(\theta_2,\xi){\cal M}(\theta_1,\xi ).
\end{equation}

Once the total $4\times 4$ transfer matrix is computed, one may find 
the transmission amplitudes $T_j(N)$ and reflection amplitudes $R_j(N)$ 
for the two polarization directions $(j=1,2)$ by solving the linear 
algebra problem 
\begin{equation}
{\cal M}_{tot,N}
\pmatrix{1+R_1(N) \cr R_2(N)  \cr R_2(N) \cr 1-R_1(N)}=
\pmatrix{T_1(N) \cr T_2(N)  \cr -T_2(N) \cr T_1(N)}.
\end{equation}  
For an incident photon in the allowed $1$-polarization direction, 
the probability that the photon exits a stack of $N$ polarizers 
(rotated into the allowed $2$-polarization direction) is given by
the absolute square of the transmission amplitude,
\begin{equation}
P_N=|T_2(N)|^2.
\end{equation}
From Eqs.(54), (59), and (61)-(63), with each of the angles in 
$(\theta_1,...,\theta_N)$ set equal to $(\pi/2N)$, one may compare 
the the projection postulate $P_N$ in Eq.(7) to the Maxwell theory, 
or equivalently the photon Schr\"odinger Eq.(63), without the projection 
postulate. The matrix multiplications in Eq.(61) and the linear algebra 
in Eq.(62) are best left to a computer. The results of such a computation 
are shown in Fig.3.

\begin{figure}[htbp]
\begin{center}
\mbox{\epsfig{file=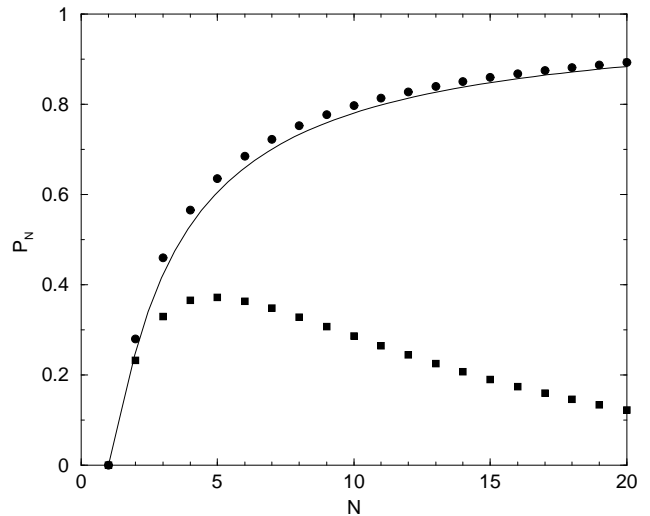,height=70mm}}
\caption{The Zeno effect probability $P_N$ for a photon to pass 
through $N$ polarizers: (i)The solid curve (evaluated for integer $N$) is 
the projection postulate in Eq.(7). (ii) The filled circles represent 
the Maxwell theory with zero attenuation for the allowed polarization 
direction. (iii) The filled squares represent the Maxwell theory with a 
small attenuation even for the allowed polarization direction.}
\label{zfig3}
\end{center}
\end{figure}

The solid curve in Fig.3 is the prediction of the projection postulate 
Eq.(7). For the Maxwell theory with a reasonable stack of laboratory 
polarizers, there may be $\sim 20$ wavelengths of light in each polarizer 
in the stack. Thus we chose the parameter $\xi=(\omega a/c)=100$ for 
purpose of illustration. We further choose  $\tilde{\epsilon}_1=1$ 
to describe the polarization direction for which the polarizer 
would be transparent and the complex $\tilde{\epsilon}_1=1+2i$ 
to describe the polarization direction for which the polarizer 
would be strongly attenuated (if not absolutely opaque). 
The resulting computed points are the filled circles of Fig.3. 
For this case, the projection postulate yields a $P_N$ which is quite close 
to the full Maxwell theory. 

For some regimes (perhaps more close to the laboratory situation), 
the Maxwell theory begins to diverge from from the projection postulate 
for large $N$. In the laboratory all polarization directions are subject 
to small (but certainly finite) attenuation. To test this situation, 
we computed the Maxwell theory prediction when the allowed 
polarization has a weak attenuation response 
$\tilde{\epsilon}_1=1.1+0.001i$ while the disallowed polarization 
has a stronger attenuation response 
$\tilde{\epsilon}_2=1.1+0.05i$. The results are the filled squares of 
Fig.3. For $N\le 3$, there is a Zeno effect in which $P_N$ increases. 
However for $N\ge 6$, the is an attenuation induced decrease in $P_N$,  
contrary to the ``Zeno effect expectation''. However, one might regard 
this attenuation of the allowed polarization direction to be merely an 
indication of poor experimental design.

Some final comments about the calculations producing Fig.3 are 
worthy of note. The transfer matrix formalism used to calculate the 
transmission probability $P_N$ was tedious, but straight forward. However, 
the numerical computations required to produce Fig.3 were not entirely 
trivial. The first three numerical commercial software programs that 
we employed for the calculation produced complete nonsense\cite{41}. 
The physical problem was that the Maxwell waves reflect back and forth 
within the polarizers. Hence, there exist {\em exponentially attenuated} 
terms as well as {\em exponentially amplified} terms in 
the transfer matrix. High accuracy was needed for the matrix 
multiplications in Eq.(61). Only with high accuracy at intermediate 
stages of the calculation, could Eqs.(62) and (63) be 
reliably computed. For realistic laboratory polarizers, the 
required accuracy turns out to be very high indeed. Fortunately, 
Aberth\cite{42} has developed precise numerical algorithms together 
with publicly available programming code which allows for $\sim 10^4$ 
significant figure accurate computations. Much  of the required code 
for the standard functions and standard linear algebra was thus available. 
We used this code. However, we required $\sim 10^3$ significant figure 
computations to produce the reliable plots shown in Fig.3. 
 
\section{Conclusions}

In discussions of the foundations of quantum mechanics there has been 
a tendency to regard all interference of waves as explicitly due 
to quantum mechanics. This is only sometimes true, since ``waves'' as well 
as ``particles'' also have classical limits. In many (perhaps most) 
of the electromagnetic experiments and applications, the Maxwell wave 
itself is classical. The {\em distinction} between a classical wave, 
a classical particle, and a quantum particle has been discussed simply, 
{\it yet clearly}, by Feynman\cite{43}. 

(i) A quantum particle is {\em certainly not} a classical particle. 
Quantum particle dynamics dictates amplitude superposition, 
whereas classical dynamics dictates probability superposition. 
(ii) A quantum particle is {\em certainly not} a classical wave. 
A classical wave is ``smooth'' while a quantum particle comes in ``lumps''. 
Photons are lumpy. Photon counters yield one click whenever a 
photon is detected. {\em One} photon can never make {\em two} different 
counters each give you {\em one half} of a click. That is much too smooth 
for quantum mechanics. In quantum mechanics, each photon counter gives 
one click or no clicks. The photon is {\it certainly not} a classical 
wave. The photon is {\em certainly not} a classical particle. The photon is 
{\em certainly not} a little bit classical wave and a little bit 
classical particle. (iii) The photon is {\em neither} a classical wave 
{\em nor} a classical particle {\em nor} a little bit of both. 

Electromagnetic waves, as described by Maxwell, are classical waves. 
If a radio station broadcasts a signal, then the signal is  
(for all practical purposes) smooth. Your radio gets 
some of the signal, and our radio gets some of the signal. 
There is enough smoothly distributed classical wave to go around 
to all of the radios. The classical Maxwell wave is smooth. One detector 
cannot get a fraction of a photon. Remember, photons come in lumps. 
One photon can get destroyed in at most one radio receiver. One 
detector can get any fraction of a classical Maxwell wave. Remember, 
classical waves are smooth. Yet the Schr\"odinger equation for photon 
is the same (in a mathematical formalism) as the Maxwell equations for  
classical electromagnetic waves. This is because the photon has no mass 
and because the photon is a Boson. When very many photons are Bose 
condensed into a single quantum state, the result is a Maxwell wave 
which is classical for all practical purposes. 

The above picture was not all that clear to Einstein (by his own admission), 
and we do not claim to know more about photons than did Einstein. 
According to Einstein, a Maxwell wave hits a half silvered mirror 
and some of the wave goes to one detector and the rest of the wave 
goes to a second detector. If the wave represents one photon, then 
only one counter goes click. The other counter gets whacked with a 
Maxwell wave. Yet this counter counts nothing. When there is but one 
photon, only one counter can speak for her. Some Maxwell wave 
parts turn out to have have a photon, and some Maxwell wave parts 
turn out not have a photon. Some counters go click and some counters 
remain mute. You never know. And photons are lumpy. It is a very strange 
shell game according to Einstein\cite{44}.

However, the Zeno effect formally attributed to the quantum mechanical 
projection postulate has here been argued to be merely a classical wave 
effect. Maxwell would not have had to increase his knowledge of 
electromagnetic theory by even one iota in order to totally 
understand the polarization version of the Zeno effect. No quantum 
mechanics is required, not to even speak of the projection postulate. 
    
\vfill \eject

\medskip
\centerline{\bf Acknowledgement}
\medskip
\par \noindent
This work has benefited from the allocation of time at the Northeastern 
University High-Performance Computing Center (NU-HPCC).

\end{document}